# Navigating the EU AI Act: Foreseeable Challenges in Qualifying Deep Learning-Based Automated Inspections of Class III Medical Devices

Julio Zanon Diaz[1*], Tommy Brennan[2†] and Peter Corcoran[1†]

[1*]School of Electrical and Electronic Engineering, University Of Galway, University Road, Galway, H91 TK33, Ireland.
[2]Visual-Cognitive Manufacturing Group, Digital Manufacturing Ireland, Castletroy, V94 237R, Co. Limerick, Ireland.

*Corresponding author(s). E-mail(s): J.ZanonDiaz1@UniversityOfGalway.ie;
Contributing authors: Tommy.Brennan@DMIreland.org; Peter.Corcoran@UniversityOfGalway.ie;
†These authors contributed equally to this work.

**Abstract**

As deep learning (DL) technologies advance, their application in automated visual inspection for Class III medical devices offers significant potential to enhance quality assurance and reduce human error. However, the adoption of such AI-based systems introduces new regulatory complexities—particularly under the EU Artificial Intelligence (AI) Act, which imposes high-risk system obligations that differ in scope and depth from established regulatory frameworks such as the Medical Device Regulation (MDR) and the U.S. FDA Quality System Regulation (QSR). This paper presents a high-level technical assessment of the foreseeable challenges that manufacturers are likely to encounter when qualifying DL-based automated inspections—specifically static models—within the existing medical device compliance landscape. It examines divergences in risk management principles, dataset governance, model validation, explainability requirements, and post-deployment monitoring obligations. The discussion also explores potential implementation strategies and highlights areas of uncertainty, including data retention burdens, global compliance implications, and the practical difficulties of achieving statistical significance in validation with limited defect data. Disclaimer: This paper presents a technical perspective and does not constitute legal or regulatory advice.

# 1 Automated Visual Inspection of Class III Medical Devices

The commercialisation and manufacture of medical devices are subject to regulation by various authorities, each corresponding to a specific geographical jurisdiction. These regulatory bodies implement distinct frameworks that generally include a risk-based classification system, grouping devices by product family or level of risk. Devices assigned to the highest classification tier are subject to the most stringent regulatory oversight. Typically, these include devices that are life-sustaining, life-supporting, implanted, or those that pose a significant risk of illness or injury if they fail. Examples include implantable stents, pacemakers, heart valves, and deep brain stimulators. Table A1 in Annex A presents a summary of the principal regulatory authorities and the corresponding medical device classification systems employed by each.

This paper will concentrate exclusively on devices within the highest risk category, as these present the most pronounced challenges in qualification and regulatory compliance. Despite notable regional differences in regulations governing both commercialisation and manufacturing, requirements concerning manufacturing practices—particularly those related to visual inspection—are closely harmonised.

**Disclaimer:** This publication is intended solely as an academic and technical evaluation. It is not a substitute for legal advice or official regulatory interpretation. The information presented here should not be relied upon to demonstrate compliance with the EU AI Act or any other statutory obligation. Manufacturers are encouraged to consult appropriate regulatory authorities and legal experts to determine specific compliance pathways.

In line with current regulatory expectations under both the EU Medical Device Regulation (MDR) and the U.S. FDA Quality System Regulation (QSR), this assessment focuses exclusively on static deep learning models—that is, models which are trained prior to deployment and not subject to continual or periodic retraining in production. This reflects existing validation obligations for manufacturing software, which require that software tools affecting product quality be fully validated prior to use and revalidated following any changes [1] [2]. As such, adaptive or retrainable models would fall outside the scope of current best practices and are not considered in this analysis. For the purposes of this analysis, the focus will be on Class III devices as defined by European Commission under the Medical Device Regulation (MDR), specifically Article 51 and Annex VIII – Classification Rules[3], and by the United States Food and Drug Administration (FDA) under 21 CFR Part 860 – Medical Device Classification Procedures [4].

Visual inspection is a widely employed method of quality assurance in the manufacturing of Class III medical devices. These inspections continue to play a critical role in ensuring product quality and are predominantly performed by human operators, owing to their unique capabilities and contextual judgement, as noted by Charles et al. [5]. For conciseness, we will refer to human-conducted visual inspections as Human-Visual Inspections (HVI), and those performed using computerised logic as Automated-Visual Inspections (AVI).

Manufacturers in this sector typically implement multiple inspection stages to verify compliance with strict specifications. Rodriguez-Perez [6] underscores the industry's reliance on such inspections and highlights a significant prevalence of human error—estimated to contribute to approximately one-third of all non-conformances in medical device and pharmaceutical manufacturing. Consequently, over the past decade, there has been a marked trend toward investing in AVIs through the application of conventional computer vision technologies. However, these systems face challenges when dealing with complex inspection tasks, particularly due to their reliance on precisely defined, objective defect criteria. In contrast, HVIs often operates on subjective criteria that are readily interpreted by humans but difficult to encode into software.

The emergence of Deep Learning (DL), particularly through the use of Convolutional Neural Networks (CNNs) for image analysis, has enabled engineers to explore the automation of inspections even when subjectivity is involved. These technologies can learn from labelled image data, previously annotated by human experts, thereby bridging the gap between subjective human judgement and machine-based analysis. While this development represents a significant opportunity for innovation in manufacturing, it also introduces new challenges—chiefly, the absence of comprehensive regulatory guidance for the qualification and deployment of DL-based inspection systems within the medical device manufacturing framework.

## 2 Applicable Medical Device Regulatory Framework

This chapter explores the principal regulations and standards applicable to automated inspection systems employed in the manufacture of Class III medical devices, with particular reference to the frameworks established by European and United States regulatory authorities. Within the European Union, the principal legislative instrument governing the manufacture of medical devices is the Medical Device Regulation (MDR) (EU) 2017/745. For the purposes of a focused and relevant discussion, special cases such as drug–device combinations, where Class III implantable devices are used to deliver medicinal substances—for example, drug-eluting stents (DES)—will be excluded. In such instances, while the device remains subject to the Medical Device Regulation (MDR), a scientific consultation with the European Medicines Agency (EMA) or a national medicines authority may be required regarding the medicinal component. Compliance with the MDR is generally achieved through adherence to designated harmonised standards, which serve as a presumption of conformity. Among the most relevant harmonised standards for manufacturing are:

- **EN ISO 13485:2016+A11:2021** – *Quality management systems – Requirements for regulatory purposes* [1].

- **EN ISO 14971:2019+A11:2021** – *Application of risk management to medical devices* [7].

It is also important to acknowledge the existence of standards tailored to Software as a Medical Device (SaMD), some of which address specific requirements for machine learning applications. These standards however are not applicable for software used to manufacture Medical Devices and therefore has not been included in the assessment. In the United States, the primary regulatory framework for medical device manufacture is established by the Food and Drug Administration (FDA) and includes:

- **21 CFR Part 820** – *Quality System Regulation (QSR)* [2].
- **21 CFR Part 11** – *Electronic Records and Signatures* [8].

Notably, the FDA recognises ISO 13485 and ISO 14971 as acceptable standards for demonstrating regulatory compliance, and significant efforts are underway to formally harmonise ISO 13485 within the forthcoming Quality Management System Regulation (QMSR) framework [9]. Beyond these legally binding regulatory requirements, specific to medical devices, several generic quality standard and non-harmonised but widely accepted industry guidelines are often used by manufacturers to strengthen internal quality procedures. Although these are not covered in detail within this assessment, it is useful to acknowledge the most commonly referenced:

- **ISO 9000:2015** – *Quality Management Systems – Fundamentals and vocabulary* [10].
- **ISO/TR 80002-2:2017** – *Guidance on validating software used in quality systems (non-product software)* [11].
- **ISPE GAMP 5** – *Good Automated Manufacturing Practice* [12].
- **GHTF SG3/N99-10:2004** – *Quality Management Systems – Process Validation Guidance* [13].
- **FDA Guidance** – *Process Validation: General Principles and Practices (2011)* [14].

## 2.1 Quality Management System and Risk Management

Automated inspection systems must be implemented within a documented and validated Quality Management System (QMS) and governed by a risk-based approach. The combined expectations include:

- *[ISO 13485:2016* [1]*, Clauses 4.1, 4.2.4, 4.2.5, 7.1, 8.1*] A QMS must be established that is tailored to medical devices, incorporating document and record control, traceability, change control, and regulatory compliance.
- *[ISO 13485:2016* [1]*, Clauses 7.5.1, 7.5.6; 21 CFR §820.70, §820.75* [2]] Inspection processes (manual or automated) must be defined using pre-approved, objective, and risk-aligned acceptance criteria, and validated if their outputs cannot be verified through subsequent inspection.
- *[ISO 13485:2016* [1]*, Clause 4.1.6; 21 CFR §820.70(i)* [2]*; 21 CFR §11.10(a)* [8]] Inspection software (e.g., machine vision) must be validated for its intended use, with documented protocols, verification results, and version control.

- *[ISO 13485:2016* [1]*, Clause 7.6; 21 CFR §820.72* [2]] Inspection equipment (e.g., image acquisition apparatus, measurement tools) must be qualified, calibrated, and maintained to ensure accuracy and suitability.
- *[ISO 14971:2019* [7]*, Clause 7, 7.2, 7.3, 10*] Inspection processes must be embedded in the device risk management strategy, with documented identification of potential failure modes, including:
  - Environmental factors (e.g., lighting),
  - Operator error (for semi-automated systems),
  - Software limitations (e.g., misclassification).
- *[ISO 14971:2019* [7]*, Clause 7.3, Clause 10*] Effectiveness of risk control measures must be verified and monitored for residual risk throughout production and post-market phases.

## 2.2 Data Integrity and Inspection Results

Where inspection systems generate or rely on electronic records, full compliance with data integrity, traceability, and security controls is mandatory:

- *[21 CFR §11.10(b)-(k)* [8]] Electronic records, including pass/fail outcomes and batch disposition data, must be:
  - Secure, retrievable, and human/electronically readable,
  - Audit-trailed with timestamps, operator identity, and record of parameter changes,
  - Protected against unauthorised access via role-based permissions.
- *[21 CFR §11.10(f),(g)* [8]*; ISO 13485:2016* [1]*, Clause 6.2*] Access control and device checks must be enforced, and operators must be trained on inspection system operation and limitations.
- *[21 CFR §11.10(j)-(k)* [8]] Documentation control procedures must manage system updates, change history, and user accountability.
- *[ISO 13485:2016* [1]*, Clauses 7.5.1, 8.2.6; 21 CFR §820.80, §820.181* [2]] Inspection outcomes must be recorded and linked to product identifiers in the Device History Record (DHR), retained for regulatory inspection.
- *[ISO 13485:2016* [1]*, Clause 8.3; 21 CFR §820.100* [2]] Failed inspection results must be subject to nonconformance handling and root cause analysis as part of the Corrective and Preventive Action (CAPA) system.
- *[ISO 13485:2016* [1]*, Clause 4.1.6; 21 CFR §820.70(i)* [2]*; 21 CFR §11.10(a)* [8]] Any modification to inspection software, acceptance limits, or configuration workflows must follow change control and revalidation protocols.

## 2.3 Qualification Activities

- *[GAMP 5, Appendix D*] Installation Qualification (IQ) – Verifies correct system setup, including:
  - Hardware (e.g., camera models, GPU types),

– Firmware (e.g., driver/firmware versions),
– Calibration validation (e.g., pixel-to-millimetre transformations),
– Software code traceability (e.g., script version used for build).

- [GAMP 5] Operational Qualification (OQ) – Confirms functionality under expected operating conditions, verifying that:

– Inputs generate valid outputs,
– Inspection logic performs within defined ranges,
– Records comply with Part 11 or ISO 13485 data requirements.

- *[GAMP 5; ISO 13485:2016* [1]*, Clause 7.5.6*] Performance Qualification (PQ) – Also referred to as process validation, Ensures repeatability and consistency in production, typically involving Test Method Validation (TMV) with variable and/or attribute data under normal use conditions.
- *[GHTF SG3/N99-10:2004, Clause 6.3*] Sampling plans used in PQ and TMV must be statistically justified, particularly where inspection is a critical process.
- [Taylor [15]] Risk, Confidence, Reliability: Organisations should classify inspection risks based on severity and detection likelihood. Sample sizes for validation should be calculated using statistical methods. The Success-Run Theorem is an example of a widely accepted method for calculating sample sizes for process validation of attribute inspection methods. This approach assumes no false negatives are permitted during the validation study. There are alternative statistical methods that allow for a defined number of false negatives, these typically require significantly larger sample sizes.

$$\eta = \left\lceil \frac{\ln\left(1 - Confindence\right)}{\ln\left(Reliability\right)} \right\rceil \quad (1)$$

where,
**Confidence** is the statistical likelihood that the validation conclusion is correct. In TMV, it reflects how sure we are that the test method performs as intended, based on the validation data. For example, if the confidence level is 95%, then if the validation were repeated many times, 95% of those validations would correctly confirm the method's performance.
**Reliability** is the probability that the test method will consistently detect a defect or produce a correct result under defined conditions. For example, if Reliability is 99%, then the test method will correctly detect the defect (or give the correct result) 99 times out of 100 in routine use.
Typical values for High-risk inspections are Confidence= 95% and Reliability = 99%.

- [AIAG MSA [16]] – TMV Sample Composition- Best Practices: TMV must include a representative mix of defective and non-defective units, ideally:

– 25–50% defective,
– A mix of marginal cases and clearly nonconforming items,
– Artificial defects if real samples are unavailable.

# 3 AI Regulatory Framework

## 3.1 EU AI Act

The EU Artificial Intelligence (AI) Act [17], which partially entered into force on 2 August 2024, will become applicable after a transitional period of two to three years, with the majority of its provisions taking effect on 2 August 2026, as set out in Article 113. However, the full enforcement of the Act is phased. Notably, provisions related to high-risk AI systems, as defined under Article 6(1), will not come into effect until 2 August 2027. For AI systems that were already placed on the market or put into service prior to 2 August 2026, the obligations of the Act will only apply if those systems undergo significant design modifications after this date, in accordance with Article 111(2).

### 3.1.1 Harmonised Standards

To support the effective implementation of the AI Act, the European Commission has mandated the development of harmonised standards under the responsibility of CEN-CENELEC Joint Technical Committee 21 (JTC 21) [18]. These standards aim to provide technical specifications that facilitate compliance with the Act. The committee's work is focused on the following areas of the legislation:

- **AI Trustworthiness Framework**: Defining criteria for the development of trustworthy AI systems; standards are currently under development.
- **AI Risk Management**: Addressing operational risks inherent to AI systems; standards are currently under development.
- **AI Quality Management System**: Supporting the establishment of structured and consistent AI development processes; development has not yet commenced at the time of this assessment.
- **AI Conformity Assessment**: Establishing procedures for the verification of compliance; development has not yet commenced at the time of this assessment.
- **Data Governance and Quality**: Developing standards for the responsible management and high quality of data used in AI systems; development has not yet commenced at the time of this assessment.
- **Record-Keeping and Logging**: Ensuring comprehensive documentation of AI system operations to support transparency and accountability; standards are currently under development.
- **Transparency and Information Provision**: Standardising the information that should be provided to users and stakeholders; standards are currently under development.
- **Human Oversight**: Defining principles for effective human control and intervention in AI system operations; standards are currently under development.
- **Accuracy, Robustness, and Cybersecurity**: Setting out performance and security requirements for AI systems; standards are currently under development.
- **Sector-Specific Standards**: Addressing domain-specific needs such as those for machine vision applications; development has not yet commenced at the time of this assessment.

The following standards have already been published and are expected to be harmonised with the AI Act [19] [18]:

- ISO/IEC 22989:2022 – *Information technology – Artificial intelligence – Concepts and terminology* [20].
- ISO/IEC 23053:2022 – *Framework for AI systems using machine learning* [21].
- ISO/IEC 23894:2024 – *Guidance on risk management for AI systems* [22].
- ISO/IEC 27001:2013 – *Information security management systems, supporting both AI management and risk management frameworks* [23].
- ISO/IEC 5259 Parts 1–4:2025 – *Data quality for analytics and machine learning* [24], [25], [26], [27].
- ISO/IEC 42001:2023 – *Artificial Intelligence Management System* [28].

Although no explicit declarations have been made confirming harmonisation for the following standards, they have been referred by JTC 21 and are therefore expected to inform future standardisation activities in support of the AI Act [18]:

- CEN/CLC/TR 18115:2024 – *Data governance and quality for AI in the European context* [29].
- CEN/CLC ISO/IEC/TR 24027:2023 – *Bias in AI systems and AI-aided decision-making* [30].
- CEN/CLC ISO/IEC/TR 24029-1:2023 – *Assessment of the robustness of neural networks – Part 1: Overview* [31].
- EN ISO/IEC 8183:2024 – *Data lifecycle framework for AI* [32].
- CEN/CLC/TR 17894:2024 – *AI Conformity Assessment* [33].
- EN ISO/IEC 25059:2024 – *Quality model for AI systems* [34].
- CEN/CLC ISO/IEC/TS 12791:2024 – *Treatment of unwanted bias in classification and regression machine learning tasks* [35].

In Annex A, Table A2 we provide a list of JTC 21 projects that are either in the planning stage or currently in progress to complete the set of harmonised standards required under the AI Act. While this list is subject to change as the Act approaches full enforcement, it is worth noting that one such planned standard pertains specifically to computer vision: JT021025 – *Artificial Intelligence – Evaluation Methods for Accurate Computer Vision Systems*. This standard seeks to establish methodologies for assessing the accuracy of computer vision systems, thereby supporting compliance with the performance requirements of the EU AI Act.

### 3.1.2 Scope and Assumptions of the Assessment

To assess the regulatory requirements under the EU Artificial Intelligence (AI) Act pertaining to the use of Deep Learning (DL) for automated defect inspections in Class III medical device manufacturing, it is necessary to define a specific use case and make several simplifying assumptions. These assumptions establish the boundaries of the analysis presented in this paper:

1. High-Risk Classification Assumption: Although not all defects inspected by automated systems pose high-risk failure modes, this paper assumes that each automated inspection includes at least one defect associated with a high-risk condition. Medical devices classified under Class III fall within the scope of Union harmonisation legislation, specifically Regulation (EU) 2017/745 on medical devices (MDR), which is referenced in Annex II of the AI Act. Consequently, any AI system that functions as a safety component in the manufacturing process of such devices meets the criteria for high-risk classification under Article 6(1) of the AI Act. It is acknowledged that this is an inferred classification, as neither the MDR nor current harmonised standards under the AI Act explicitly reference automated inspections. Nonetheless, it is reasonable to anticipate that future revisions of these standards will formally clarify the regulatory status of such systems.
2. Model Lifecycle Assumption: It is assumed that automated inspection systems are subject to rigorous pre-deployment validation, both at the software and process level, as is standard practice in medical device manufacturing. Accordingly, DL models used in this context are treated as "static" systems—that is, models that do not undergo continual or periodic retraining once deployed into production. In the event that retraining becomes necessary, a comprehensive revalidation process would be required in accordance with regulatory expectations.
3. Application Domain Assumption: Automated inspection systems are often used for both variable inspections (assessing measurable parameters) and attribute inspections (assessing pass/fail characteristics). In this paper, it is assumed that traditional computer vision techniques are generally preferred for variable inspections due to their calibration capabilities and alignment with recognised metrological standards. In contrast, DL models are assumed to be used primarily for attribute inspections, where the system learns to recognise features or patterns indicative of nonconformities. In some cases, these DL-derived classifications may serve as input to traditional vision algorithms, which then compute dimensional outputs.
4. Standards Coverage Assumption: The overview of regulatory and technical requirements presented in this assessment is based on the currently available versions of harmonised and supporting standards, including those under the EU AI Act and general quality and risk management frameworks. In the absence of specific guidance or updates to vertical standards, such as those under the Medical Device Regulation (MDR), which may eventually be revised to align more explicitly with the AI Act in the context of manufacturing, it is assumed that existing standards provide sufficient coverage of the core requirements relevant to automated inspections for the purposes of this analysis. Future iterations of MDR-related standards are expected to offer greater clarity on compliance pathways for AI-enabled manufacturing systems, but until such guidance is formally published, this assessment proceeds on the basis that the current corpus of standards reflects the applicable regulatory expectations.

## 3.2 Quality Management System and Risk Management

High-risk AI systems—such as DL-based visual inspection tools—must operate under a documented AI Management System (AIMS), incorporate full lifecycle risk governance, and fulfil obligations under the EU AI Act and applicable harmonised standards.

- *[EU AI Act, Art. 9* [19]*; ISO/IEC 42001:2023* [28]*, Clauses 4–6*] A formal AIMS must define objectives, roles, risk tolerances, and compliance procedures across the AI lifecycle. This includes change control, ethical risk appraisal, and linkage to QMS-level activities in ISO 13485.
- *[ISO/IEC 23894:2023* [22]*, Clauses 5.2–5.4*] AI risk management must explicitly address model-specific hazards, such as:
  - Dataset shift or concept drift,
  - Adversarial vulnerability,
  - Overfitting to test datasets,
  - Automation bias or unwarranted confidence in model outputs.
- *[CEN/CLC ISO/IEC/TS 12791:2024* [35]*, Clause 7* ] Systems must be evaluated for social, ethical, and safety impacts, especially when involved in product acceptance decisions. Human oversight must be designed and documented.
- *[ISO/IEC 22989:2022* [20]*, Clause 4.3; CEN/CLC ISO/IEC/TR 24027:2023* [30]*, Clause 6*] The system must document how it mitigates dataset bias and ensures fairness, particularly for defect variability across product variants or manufacturing sites.
- *[EN ISO/IEC 8183:2024* [32]*, Clause 5.2*] The AIMS should incorporate a data lifecycle strategy, ensuring that governance, risk control, and quality assurance are maintained from data acquisition to model decommissioning.

## 3.3 Data Integrity and Inspection Results

Data used by High-Risk AI systems must meet curation, protection, integrity, and auditability requirements, both during development and in production.

- *[EU AI Act, Art. 10; ISO/IEC 5259-2:2024* [25]*, Clauses 5–6*] Datasets used for training, validation, and testing must be statistically representative, relevant, accurate, and free from bias. Documentation must include data sources, preprocessing logic, and exclusion rules.
- *[ISO/IEC 23894:2023, Clause 6.4* [22]*; ISO/IEC 42001:2023* [28]*, Clause 8.1.2*] Data privacy must be preserved during acquisition and use, especially for image data that could reveal traceable features. Anonymisation or synthetic data should be considered where applicable.
- *[CEN/CLC/TR 18115:2024* [29]*, Clause 5.3; ISO/IEC 27001:2013* [23]*, Annex A.9–A.12*] Training and test data must be stored securely, encrypted in transit and at rest, with access control. Retention duration should be defined, and must comply with the system's traceability and revalidation policies.

- *[EN ISO/IEC 25059:2024* [34]*, Clause 5.5*] Inspection results (e.g., defect classifications, confidence levels) must be interpretable, traceable to a model version, and evaluated for reproducibility across datasets.
- *[EN ISO/IEC 8183:2024* [32]*, Clauses 6.3 and 6.5*] A structured data lifecycle must be established to ensure that data integrity and consistency are preserved during ingestion, transformation, storage, and archival. The framework must include procedures for data versioning, audit trails, and data quality metrics at each lifecycle phase.

## 3.4 Qualification Activities and Performance Validation

High-risk AI systems must be qualified with statistical rigour, validated against performance targets, and monitored post-deployment to detect degradation or unintended behaviour.

- *[EU AI Act, Art. 15; ISO/IEC 5259-1:2022* [24]*, Clause 4.2; ISO/IEC 5259-4:2024* [27]*, Clause 6.2*] Validation must include:
  - Predefined metrics such as accuracy, precision, recall, F1-score, and false positive/negative rates,
  - Systematic testing across realistic variability conditions,
  - Detection of overfitting, such as inflated test performance without generalisation.
- *[ISO/IEC 5259-3:2024* [26]*, Clauses 5–6; ISO/IEC 23053:2022* [21]*, Clauses 6.4–6.7*] Validation protocols must be statistically sound, repeatable, and transparent. Acceptable statistical approaches include:
  - Cross-validation,
  - Stratified sampling of defect cases,
  - Application of Success-Run Theorem where binary (pass/fail) outputs are used.
- *[ISO/IEC 23894:2023* [22]*, Clause 8.3; ISO/IEC 42001:2023* [28]*, Clause 10.2*] Post-deployment monitoring requirements vary based on model adaptability:
  - Static AI systems (no retraining) require:
    - Periodic reviews,
    - Revalidation after environment/configuration changes,
    - Surveillance of drift in input data distributions.
  - Retrainable AI systems must:
    - Define criteria and approval workflows for retraining,
    - Maintain audit trails of all updated model versions,
    - Conduct revalidation after each retraining cycle.
- *[CEN/CLC/TR 17894:2024* [33]*, Clause 5.2.3*] A post-market monitoring plan must be in place, including:

- Key performance indicators (e.g. inspection yield, defect recall),
- Root-cause investigations for out-of-tolerance inspection outcomes,
- Processes to trigger model updates based on performance degradation.

- *[CEN/CLC ISO/IEC/TR 24029-1:2023* [31]*, Clauses 5.2 and 6.2*] AI robustness must be validated through stress testing and adversarial analysis. This includes simulating edge-case scenarios, applying input perturbations, and evaluating the system's tolerance to environmental variation and noise—critical in manufacturing inspection environments.

# 4 Assessing Foreseeable Challenges for Deep Learning Automated Inspections of Class III Medical Devices

This section explores the foreseeable challenges in adopting the EU Artificial Intelligence (AI) Act for qualifying deep learning (DL)-based automated inspections of Class III medical devices, specifically for companies that already comply with existing regulatory frameworks such as the Medical Device Regulation (MDR) and the U.S. FDA Regulation. Rather than providing a comprehensive list of new requirements—which has been addressed in earlier sections—this analysis focuses on areas of implementation that may prove difficult to align with existing practices. It draws on findings from the literature as well as expert consultation. Given that harmonised standards under the AI Act are still evolving and that the obligations for high-risk systems are not yet fully in force, there remains some degree of subjectivity in this analysis, which is expected to be clarified in future guidance and revisions.

## 4.1 Quality Management System Requirements

Many requirements related to establishing an AI Management System (AIMS)—including AI-specific risk management—overlap with existing QMS practices under MDR and FDA. However, the two frameworks are grounded in distinct conceptual foundations. For example, EN ISO 14971:2019 (MDR) is derived from ISO Guide 51:2014 [36] and centres on product safety and engineering risks. In contrast, ISO/IEC 23894:2024 is built on ISO Guide 73:2009 [37], reflecting a broader enterprise risk management approach.

Though these frameworks are not incompatible, their differing scopes and terminologies may pose challenges for alignment. It is likely that EN ISO 14971:2019, ISO 13485:2016, or both will require revision to better harmonise with AI-specific risk frameworks such as ISO/IEC 23894:2024.

In addition, there is a fundamental shift in the requirement to define roles, responsibilities, and policies across the entire AI lifecycle—from development through to deployment and retirement. In manufacturing, this presents a challenge: policies cannot be confined to inspection use cases alone, but must also account for other applications (e.g. large language models used in documentation, causal inference for root cause analysis, or DL-based supply chain prediction tools).

While smaller organisations may scale policies incrementally, larger enterprises with numerous concurrent initiatives may struggle to develop AIMS documentation (e.g., SOPs, work instructions) that is sufficiently comprehensive and future-proof.

## 4.2 Image curation: Ethical Considerations and Privacy

Images used for DL-based inspection systems typically do not pose privacy risks, especially when generated in controlled manufacturing environments without identifiable human content. However, it is possible for background footage to unintentionally include workers, particularly in constrained physical layouts. While GDPR [38] already addresses this issue for EU-based manufacturers, compliance with the AI Act's privacy provisions is not expected to introduce additional burdens in this context. However, for manufacturers operating under U.S. state-level privacy laws—which offer varied and less prescriptive coverage for workplace surveillance—additional safeguards may be required to ensure AI Act alignment.

Emerging applications where AI is used to monitor worker activity (e.g. to confirm safety-critical steps in a procedure) may trigger more complex ethical considerations. However, as this use case extends beyond product inspection, it is excluded from this assessment.

## 4.3 Image Curation: Bias and Fairness

Under MDR and FDA frameworks, bias and fairness are indirectly addressed through Test Method Validation (TMV) supported by statistically sound sampling plans. These methods are sufficient for rule-based computer vision, where functions are explicit and interpretable. For instance, in a rule-based traditional approach, verifying that a function correctly classifies samples based on the average brightness of a define region of pixels, requires only a small sample of boundary cases.

However, Deep Learning models operate on latent features and cannot be dissected in the same way. Their ability to generalise post-deployment depends entirely on training data representativeness. Ensuring robust performance therefore requires training and validation datasets that are statistically significant and reflective of real-world variability.

This represents a notable departure from current practice, particularly in the medical device sector where defect images are rare and difficult to simulate. The EU AI Act introduces rigorous expectations to mitigate bias across the system lifecycle—from training and testing to TMV and post-deployment monitoring.

Meeting these expectations will require manufacturers to establish new processes capable of demonstrating that their data is both representative and balanced.

## 4.4 Image Storage for Training and Inference Data

Emerging standards (e.g. EN ISO/IEC 8183:2024, CEN/CLC/TR 18115:2024) imply that training datasets and associated metadata must be retained to ensure traceability, auditability, and revalidation capability. Under MDR and FDA, manufacturers already retain software artefacts, model documentation, and validation records—often for periods aligned with patient life expectancy (i.e., 70+ years). However, these regulations do not require the retention of raw training images.

This presents a significant challenge as typically training and validation images are high-resolution, and datasets are large. The cost of secure long-term storage as quality records can be prohibitive, particularly at scale. Manufacturers with thousands of inspection points may find that compliance with this requirement becomes a barrier to full AI adoption in visual inspection.

## 4.5 System Output and Explainability

While the AI Act does not explicitly require the storage of inferencing images, it does require that systems support traceability and retrospective decision analysis. For inspection systems, this may necessitate storing inference-stage images alongside model decisions to enable review and audit.

Moreover, harmonised standards (e.g. EN ISO/IEC 25059:2024) recommend the use of explainability tools, such as saliency maps or surrogate models. Although framed as a ”should” rather than a ”shall”, in the context of high-risk applications, this requirement is likely to be interpreted as effectively mandatory.

At present, there is limited consensus on how explainability tools can be validated for use in regulated quality systems. Techniques such as saliency maps, class activation maps, LIME, and SHAP are widely discussed in academic literature but are highly model- and task-dependent, often rely on intuitive visual cues, and lack standardised benchmarks or evaluation criteria. As such, their interpretability is typically subjective and difficult to reproduce. Consequently, these tools may be useful for investigational purposes or root-cause analysis but are not yet suitable as formal quality assurance mechanisms in high-risk medical device manufacturing. A more robust and auditable approach, aligned with current quality system practices, is to establish traceability and control over the full lifecycle of data and model development. This includes:

- A documented process for image and defect sample acquisition that ensures statistical representation of expected variation;.
- Controlled labelling workflows that document inter-annotator variability and define pass/fail criteria objectively;
- Storage and traceability of all training and testing datasets, including metadata and version control, so they can be inspected as part of quality investigations;
- Monitoring feature distributions at inference time to detect concept drift or anomalous behaviour in deployed models.

Collectively, these practices create a data-centric transparency framework that, while not offering direct interpretability of model internals, achieves the intent behind explainability—ensuring that the relationship between inputs and outputs is controlled, verifiable, and auditable. This mirrors established process validation strategies for black-box manufacturing processes, where input control, boundary definition, and output reproducibility are prioritised over internal transparency. While academic research may continue to produce model-specific explainability techniques, such tools are unlikely to meet the generalisation, robustness, and auditability standards required in a regulated quality environment in the near term. Until then, a lifecycle-driven data and model governance strategy offers a more practical and regulation-aligned alternative.

In this context, it is noteworthy that the FDA's draft guidance on the use of artificial intelligence to support regulatory decision-making [39] outlines a structured documentation framework aimed at enhancing model interpretability, with no mention of explainability tools. Instead, it emphasises the importance of thoroughly detailing the development process of the models, elucidating how they generate conclusions, reporting performance metrics accompanied by confidence intervals, and disclosing known limitations, including potential sources of bias. It should be observed that this guidance document had not been finalised at the time of writing and is not intended to serve as a harmonised standard under the EU AI Act.

## 4.6 Software and Process Validation: Sampling and Performance Requirements

While AI validation may imply the need for larger validation datasets than traditional computer vision systems, the AI Act stops short of defining quantitative thresholds. Instead, it requires that training and validation data be of sufficient quality and representative of operational conditions.

This requirement is compatible with MDR and FDA TMV practices. For example, high-risk inspection processes can be validated using the Success-Run Theorem, setting confidence and reliability thresholds (e.g., 95% and 99%, respectively). Manufacturers are therefore unlikely to face philosophical contradictions but may require greater test volumes to compensate for Deep Learning opacity and performance variability.

## 4.7 Model Monitoring After Deployment

This assessment focuses on static AI models—those that are not retrained post-deployment. While these models offer stability, they remain subject to performance drift due to shifts in input distributions or environmental factors.

The AI Act and associated standards (ISO/IEC 23894:2023, ISO/IEC 42001:2023) do not explicitly differentiate static vs. retrainable models. Instead, they impose universal monitoring requirements, including:

- Regular performance reviews,
- Revalidation following any significant changes,
- Mechanisms for detecting data drift.

This represents a novel requirement for manufacturers. Under MDR and FDA frameworks, routine verification typically applies only to calibration of variable inspection tools. Attribute-based inspections—such as DL—driven defect classification—are generally exempt from revalidation unless a non-conformance is detected.

Therefore, aligning with the AI Act will necessitate new monitoring procedures, tools for tracking model stability, and clear triggers for revalidation—even in the absence of active retraining.

## 4.8 Current Industry Practice and Limitations

While machine learning—driven inspections remain in the early stages of adoption within the medical device industry, current deployments are generally limited and governed by existing software validation and process qualification frameworks. As outlined in GAMP 5[36] Appendix D11, these systems are validated using traditional waterfall or V—model methodologies, with models treated as locked, static "black boxes". Their lifecycle documentation typically resides within engineering documentation rather than formal QMS artefacts, as there is no explicit regulatory requirement for model transparency or retraining governance. Furthermore, post-deployment monitoring is not mandated, and practices for image storage vary. Due to the high cost of long-term storage, images used for training or inference may or may not be retained, thereby limiting the feasibility of comprehensive bias tracking or explainability analysis in practice.

# 5 Conclusions and Strategic Recommendations

## 5.1 Discussing Challenges

While Section 4 provided a detailed analysis of the foreseeable challenges imposed by the EU AI Act, this section offers a summarised, interpretive discussion focused on areas where current MDR and FDA frameworks may fall short or require adaptation.

- **Data Representativeness and Bias Mitigation**: Although traditional test method validation (TMV) under MDR and FDA includes statistically justified sampling, it does not explicitly address bias detection or diversity in AI datasets. Future AI—specific standards are expected to formalise requirements around data balance, annotation consistency, and defect representativeness.
- **Data Retention Limitations**: Current regulations emphasise long-term retention of validation records, but do not mandate retention of raw image datasets. Given the AI Act's emphasis on traceability and auditability of training data, storage strategies—such as triage—based image retention or feature-level logging—may become necessary.
- **Explainability Requirements**: The paper acknowledges that explainability tools are not yet mature enough to support regulatory use. Instead, lifecycle traceability of input data, labelling methods, and model outputs is proposed as a practical alternative, aligning with both GAMP 5 and ISO 13485 expectations.
- **Monitoring of Static Models**: MDR and FDA do not currently mandate performance surveillance for non-adaptive systems. However, the AI Act introduces continuous monitoring obligations even for locked models. This highlights the need for system-wide monitoring frameworks capable of detecting input drift and performance anomalies post-deployment.
- **Global and Cross-Jurisdictional Compliance**: As AI systems embedded in manufacturing environments may impact devices exported to the EU, non-EU manufacturers will also need to consider how lifecycle traceability and documentation can demonstrate conformity under Article 2 of the AI Act.

## 5.2 Proposed Lifecycle for ML-Based Visual Inspection Systems

To bridge the regulatory gaps identified above, we propose a structured lifecycle designed specifically for **static, non-adaptive machine learning systems** used in **automated visual inspections** for Class III medical devices. This lifecycle is informed by *GAMP 5* Appendix D11, ISO 13485, and Annex 11, and includes the following stages:

1. **Lifecycle Policy Definition**
   Establish a formal AI lifecycle policy scoped to the use of locked ML models in visual inspection applications. This policy should clearly distinguish these systems from other AI applications (e.g., adaptive models, decision support systems, or models handling personal data), which may have different regulatory implications.
2. **Dataset Documentation**
   Use a Dataset Specification Sheet to ensure quality, traceability, and reproducibility of training, testing, and validation datasets. The documentation should include:
   - Dataset structure, versioning, and storage location
   - Class distributions and sample quantities per class
   - Image acquisition setup and parameters
   - Defect generation or sampling methodology
   - Labelling types and process, including annotator instructions
   - Sources of bias and mitigation strategies
3. **Model Development and Evaluation**
   Employ metrics appropriate for imbalanced classification tasks, such as:
   - Balanced accuracy
   - Precision and recall
   - Escape rate
   - Confusion matrix analysis

   Ensure strict separation between training, validation, and test sets. Final test datasets must be locked and never used during model development or architecture selection to prevent bias and overfitting.
4. **Prototype Deployment Phase**
   Introduce a pre-qualification observation step, where the model is deployed in production with hidden outputs. Manual inspections remain active during this phase. The model runs passively for a statistically defined period (e.g., 2-3 months) to collect outcome data. Results are retrospectively analysed and compared with human inspections. Progression to validation only occurs if performance aligns with expectations.
5. **Test Method Validation (TMV)**
   Conduct full TMV according to established regulatory practices. Apply statistically justified sampling strategies (e.g., the Success-Run Theorem) to demonstrate sensitivity, reliability, and reproducibility of the model under production conditions.

6. **Post-Deployment Monitoring**
   Implement lightweight monitoring frameworks to detect performance drift. Recommended practices include:

   - Designing models to output intermediate feature vectors to enable ongoing distribution analysis
   - Centralising monitoring across stations using shared cloud infrastructure
   - Employing triage-based image retention (e.g., storing all defect images and a random sample of non-defect images) to manage storage burden while maintaining traceability

This proposed lifecycle integrates both current best practices in software validation and emerging requirements under the EU AI Act. It offers a scalable and regulation-aligned pathway for the deployment of machine learning in regulated inspection environments.

## 5.3 Closing Reflections and Future Work

The proposed lifecycle strategy outlined in this paper offers a practical and risk-based pathway for aligning ML-based visual inspection systems with both current regulatory frameworks and emerging obligations under the EU AI Act. By focusing on locked models, data lifecycle traceability, and input/output reproducibility, the approach remains consistent with ISO 13485, FDA QSR, and the principles established in GAMP 5, while anticipating future requirements for transparency, bias mitigation, and post-deployment monitoring.

This lifecycle addresses many of the technical and procedural gaps identified in Section 4 and summarised in Appendix B, Table B3. It offers a feasible implementation pathway that builds on standardised validation practices such as Test Method Validation (TMV) and software change control, while introducing new concepts such as the blinded prototype phase and feature-based monitoring.

Despite these advances, several critical areas remain unresolved and will require further development of standards, tools, or regulatory guidance:

- **Explainability:** There is currently no validated or auditable framework for applying explainability tools in quality-critical inspection systems. Practical alternatives—such as data lifecycle traceability—should continue to be explored until regulator-approved techniques emerge.
- **Bias and fairness validation:** The industry lacks consensus on metrics, thresholds, or test methods for evaluating dataset bias, particularly in rare-defect inspection scenarios.
- **Conformity assessment:** No clear conformity pathway currently exists for AI systems deployed as part of manufacturing quality infrastructure. This may limit adoption until vertical guidance becomes available.
- **Monitoring infrastructure:** Scalable post-deployment monitoring solutions remain an open challenge, especially where cloud-based feature tracking is not yet available or approved in regulated environments.

To support future readiness, we recommend prioritising the following areas of industry and regulatory collaboration:

- Establishing standardised templates for image dataset documentation (e.g., Dataset Specification Sheets)
- Defining minimal performance and monitoring requirements for locked models used in critical visual inspections
- Creating cross-functional working groups to inform the development of harmonised vertical standards for inspection-based AI under the EU AI Act
- Expanding the GAMP 5 framework to formally incorporate model lifecycle artefacts and risk controls specific to manufacturing AI

Ultimately, the success of AI-enabled inspections in regulated environments will depend on the ability of manufacturers, regulators, and standards organisations to co-develop transparent, auditable, and scalable frameworks that balance innovation with risk-based assurance.

# Declarations

**Competing interest:** The authors declare no competing interests.

# Appendix A Medical Devices Classifications and Standards

**Table A1** Medical Devices Regulatory Agencies and Classification Systems

| Geographical Area | Agency | Regulations defining Classification | Classification system (Highest Risk in Bold) |
|---|---|---|---|
| EU | European Commission (EC) | Medical Device Regulation (MDR) – Article 51, Annex VIII – Classification Rules [3] | Class I, II, **III** |
| USA | Food and Drug Administration (FDA) | 21 CFR Part 860 – Medical Device Classification Procedures [4] | Class I, II, **III** |
| UK | Medicines and Healthcare products Regulatory Agency (MHRA) | Medical Devices Regulations (UK MDR) SI 2002 No 618, PART II General Medical Devices [40] | Class I, II, **III** |
| China | National Medical Products Administration (NMPA) | Rules for Classification of Medical Devices" (Decree No. 15) Article 4 [41] | Class I, II, **III** |
| Japan | Pharmaceuticals and Medical Devices Agency (PMDA) | Japanese Medical Device Nomenclature (JMDN) [42] | Class I, II, III, **IV** |
| Canada | Health Canada | Risk-based Classification System for Non-In Vitro Diagnostic Devices (non-IVDDs) [43] | Class I, II, III, **IV** |
| Brazil | Agência Nacional de Vigilância Sanitária (ANVISA) | Collegiate Board Resolution RDC No. 751/2022 [44] | Class I, II, III, **IV** |
| South Korea | Ministry of Food and Drug Safety (MFDS) | South Korea, as detailed in the Medical Device Regulatory System [45] | Class I, II, III, **IV** |
| Australia | Therapeutic Goods Administration (TGA) | Classification of medical devices that are not IVD [46] | Class I, IIa, IIb, **III** |
| India | Central Drugs Standard Control Organization (CDSCO) | Medical Devices Rules, 2017 [47] | Class A, B, C, **D** |

International Organization for Standardization (ISO)

**Table A2** CEN/CLC/JTC 21 Work programme as published on the 11$^{th}$ of Jun 2025 [18]

| Project | Title | Status |
|---|---|---|
| prCEN/CLC/TR XXX (WI=JT021009) | AI Risks - Check List for AI Risks Management | Preliminary |
| prCEN/TS (WI=JT021034) | Guidelines on tools for handling ethical issues in AI system life cycle | Preliminary |
| prEN ISO/IEC 24970 (WI=JT021021) | Artificial intelligence — AI system logging | Under Drafting |
| EN ISO/IEC 22989:2023/prA1 (WI=JT021031) | Information technology — Artificial intelligence — Artificial intelligence concepts and terminology — Amendment 1 | Under Drafting |
| EN ISO/IEC 23053:2023/prA1 (WI=JT021032) | Framework for Artificial Intelligence (AI) Systems Using Machine Learning (ML) — Amendment 1 | Under Drafting |
| prEN XXX (WI=JT021038) | AI Conformity assessment framework | Under Drafting |
| prEN XXX (WI=JT021044) | **Artificial Intelligence - Taxonomy of AI tasks in computer vision – Taxonomy of AI system methods and capabilities** [1] | Under Drafting |
| prEN ISO/IEC 42102 (WI=JT021045) | Information technology - Artificial intelligence – Taxonomy of AI system methods and capabilities | Under Drafting |
| prEN ISO/IEC 25029 (WI=JT021046) | Artificial intelligence - AI-enhanced nudging | Under Drafting |
| FprEN ISO/IEC 12792 (WI=JT021022) | Information technology - Artificial intelligence - Transparency taxonomy of AI systems (ISO/IEC FDIS 12792:2025) | Under Approval |
| prEN XXX (WI=JT021029) | Artificial intelligence - Cybersecurity specifications for AI Systems | Under Drafting |
| prEN XXX (WI=JT021037) | Artificial Intelligence – Quality and governance of datasets in AI | Under Drafting |
| prEN ISO/IEC 23282 (WI=JT021012) | Artificial Intelligence - Evaluation methods for accurate natural language processing systems | Under Drafting |
| prEN ISO/IEC 25059 rev (WI=JT021027) | Software engineering - Systems and software Quality Requirements and Evaluation (SQuaRE) - Quality model for AI systems (ISO/IEC 25059:2023) | Under Drafting |
| (WI=JT021030) | Contributions towards ISO/IEC 27090 | Preliminary |
| prCEN/TS (WI=JT021033) | Guidance for upskilling organisations on AI ethics and social concerns | Preliminary |
| prCEN/TS (WI=JT021035) | Sustainable Artificial Intelligence – Guidelines and metrics for the environmental impact of artificial intelligence systems and services | Preliminary |
| prEN XXX (WI=JT021036) | Artificial Intelligence - Concepts, measures and requirements for managing bias in AI systems | Under Drafting |
| prEN XXX (WI=JT021039) | Artificial intelligence - Quality management system for EU AI Act regulatory purposes | Under Drafting |
| prEN XXX (WI=JT021019) | Competence requirements for professional AI ethicists | Under Drafting |
| prEN 18229 (WI=JT021008) | AI trustworthiness framework | Under Drafting |
| prCEN/CLC/TR XXX (WI=JT021026) | Impact assessment in the context of the EU Fundamenta215Rights | Preliminary |
| prEN ISO/IEC TR 23281 (WI=JT021002) | Artificial Intelligence - Overview of Al tasks and functionalities related to natural language processing | Under Drafting |
| prEN XXX (WI=JT021025) | **Artificial Intelligence – Evaluation methods for accurate computer vision systems** [1] | Under Drafting |
| prEN 18228 (WI=JT021024) | AI Risk Management | Under Drafting |

[1]Computer Vision harmonised standards under JT 21 have been proposed but not started. These standards will be highly relevant to the validity of this assessment as they may introduce additional guidelines.

# Appendix B Summary of Requirements and Interim Recommendations

**Table B3** Mapping EU AI Act Requirements to Interim Recommendations and Future Needs

| EU AI Act Requirement | Interim Recommendation | Future Work / Regulatory Watch-Outs |
|---|---|---|
| High-Risk Classification | Based on current standards we believe Visual inspections are High-Risk. | Additional Clarification is needed in harmonised standards or EU AI Act Annex II |
| AI-specific Quality Management System (AIMS) | While General QMS exists (ISO 13485), We proposed the introduction of a AI-specific policy, defining detail use-cases, as described in section 5.2(1). | Vertical AIMS standards under development (e.g., ISO/IEC 42001) |
| Lifecycle Documentation of AI Models | We propose formal traceability via a Dataset Sheet and test control procedure as described in section 5.2(2). | Additional guidance required for integrating model lifecycle into QMS artifacts |
| Dataset Governance and Bias Mitigation | While current Sampling plans practices partially address this requirement, we propose the creation of procedures to carry out dataset curation, labelling protocol and bias identification as described in section 5.2(2). | Needs future guidance on bias metrics, fairness audits, and representative sampling |
| Explainability of AI Decisions | We propose the use of a documented lifecycle as described in section 5.2. while robust explainability tools are developed. | Need for the development of auditable explainability tools |
| Performance Validation (TMV) | Requirement covered by current MDR/FDA Regulations. | N/A |
| Post-Deployment Monitoring | We proposed feature-based drift detection as detailed in section 5.2(6). | Future regulations likely to develop guidance related to monitoring minimum criteria |
| Change Management and Retraining | Requirement for static models already covered by current MDR/FDA Regulations. | N/A |
| Conformity Assessment of AI Systems | We propose the creation of an assessment based on the proposed lifecycle in section 5.2. | Assessment requirements not clear. Need for a defined conformity route (e.g., via notified bodies) under AI Act |
| Data and Image Retention | We recommend practices such as triage storage and feature logging to optimise storage cost. | Guidance required on storage scope, duration, and traceable artefact formats |